\documentclass[aps,pra,showpacs,floatfix,twocolumn]{revtex4}
\usepackage{amsmath}
\usepackage{bm}
\usepackage{dcolumn}
\usepackage{url}
\begin{document}
\title{Blackbody radiation shift in $^{87}$Rb frequency standard }
\author{M. S. Safronova}
\affiliation{Department of Physics and Astronomy,
 University of Delaware, Newark, Delaware 19716 and\\
 Joint Quantum Institute, University of Maryland Department of Physics and\\
National Institute of Standards and Technology, College Park, MD 20742}
\author{Dansha Jiang}
\affiliation{Department of Physics and Astronomy,
 University of Delaware, Newark, Delaware 19716}
\author{U. I. Safronova}  \altaffiliation[Permanent address: ]
{Institute of Spectroscopy, Russian Academy of Science, Troitsk, Moscow,
Russia}
 \affiliation {Physics Department, University of Nevada, Reno, Nevada 89557}

\date{\today}
\begin{abstract}

The operation of atomic clocks is generally carried out at room temperature,
whereas the definition of the second refers to the clock transition in an atom
at absolute zero. This implies that the clock transition frequency should be
corrected in practice for the effect of finite temperature of which the leading
contributor is the blackbody radiation (BBR) shift. Experimental measurements
of the BBR shifts are difficult. In this work, we have calculated the blackbody
radiation shift of the ground-state hyperfine microwave transition in $^{87}$Rb
using the relativistic all-order method and carried out detailed evaluation of
the accuracy of our final value. Particular care is taken to accurately account
for the contributions from highly-excited states. Our predicted value for the
Stark coefficient, $k_S=-1.240(4)\times 10^{-10}\text{Hz/(V/m)}^{2}$ is three
times more accurate than the previous calculation \cite{BBR-DF}.

 \pacs{06.30.Ft, 31.15.ac, 31.15.ap, 32.70.Cs}

\end{abstract}

\maketitle

\section{Introduction}

The present definition of the second in the International System of Units (SI)
is based on the microwave transition between the two hyperfine levels ($F = 4$
and $F = 3$) of the $^{133}$Cs ground state and refers to the clock transition
in an atom at absolute zero. The relative standard uncertainty of the Cs
microwave frequency standard is $4\times10^{-16}$ \cite{NIST-Csclock} at the
present time.  In 2006, the International Committee for Weights and Measures
(CIPM) recommended \cite{CIPM1} that the ground-state hyperfine microwave
transition in $^{87}$Rb \cite{Rb-1,Rb0} be used as secondary representation of
the second, along with four optical transition frequencies.

 The operation of atomic clocks is generally
carried out at room temperature implying that the clock transition frequency
should be corrected for effects of finite temperature, of which the leading
contributor is the blackbody radiation (BBR) shift. The BBR shift at room
temperature effecting the Cs microwave frequency standard has been calculated
to high accuracy (0.35\% and 1\%, respectively) in Refs.~\cite{beloy,angstmann}
implying a $6\times10^{-17}$ fractional uncertainty. These calculations are in
agreement with a 0.2\% measurement \cite{BBRcs-simon}.

 \begin{table*} [ht]
\caption{Selection of the ``best set'' values for the $5p_j-ns$, $6p_j-ns$, and
$7p_j-ns$ electric-dipole reduced matrix elements. See text for details.
Absolute values of the lowest-order DHF and SD all-order values in a.u. and
their relative difference in \% are given in columns 2-4. \label{taba-1} }
\begin{ruledtabular}
\begin{tabular}{lcccccccc}
\multicolumn{1}{c}{Transition}&
 \multicolumn{1}{c}{DHF}&
  \multicolumn{1}{c}{SD}&
   \multicolumn{1}{c}{$\Delta$(SD-DHF)}&
    \multicolumn{1}{c}{Final}&
     \multicolumn{1}{c}{Source}&
       \multicolumn{1}{c}{Unc. (\%)}&
\multicolumn{1}{c}{Unc. source} & \multicolumn{1}{c}{``Best set''}
     \\
 \hline
$    5p_{1/2}  -   5s   $   &   4.8189  &   4.2199  &   14.2\%  &   4.2310  &   Expt    &   0.07\%  &   Expt.   &   4.231(3)    \\
$    5p_{1/2}  -   6s   $   &   4.2564  &   4.1187  &   3.3\%   &   4.1458  &   SDsc    &   0.66\%  &     SDsc-SD   &   4.146(27)   \\
$    5p_{1/2}  -   7s   $   &   0.9809  &   0.9543  &   2.8\%   &   0.9527  &   SDsc    &   0.17\%  &    SDsc-SD    &   0.953(2)    \\
$    5p_{1/2}  -   8s   $   &   0.5139  &   0.5037  &   2.0\%   &   0.5022  &   SDsc    &   0.30\%  &    SDsc-SD    &   0.502(2)    \\
$    5p_{1/2}  -   9s   $   &   0.3380  &   0.3326  &   1.6\%   &   0.3314  &
SDsc    &   0.36\%  &    SDsc-SD    &   0.331(1)    \\ [0.3pc]
$    5p_{3/2}  -    5s      $   &   6.8017  &   5.9551  &   14.2\%  &   5.9780  &   Expt    &   0.08\%  &   Expt    &   5.978(5)    \\
$    5p_{3/2}  -    6s      $   &   6.1865  &   6.0135  &   2.9\%   &   6.0472  &   SDsc    &   0.56\%  &   SDsc-SD &   6.047(34)   \\
$    5p_{3/2}  -    7s      $   &   1.3925  &   1.3521  &   3.0\%   &   1.3497  &   SDsc    &   0.18\%  &   SDsc-SD &   1.350(2)    \\
$    5p_{3/2}  -    8s      $   &   0.7265  &   0.7098  &   2.4\%   &   0.7077  &   SDsc    &   0.29\%  &   SDsc-SD &   0.708(2)    \\
$    5p_{3/2}  -    9s      $   &   0.4771  &   0.4677  &   2.0\%   &   0.4662  &   SDsc    &   0.34\%  &   SDsc-SD &   0.466(2)    \\
[0.3pc]
$    6p_{1/2}  -   5s   $   &   0.3825  &   0.3335  &   14.7\%  &   0.3248  &   SDsc    &   2.69\%  &   SDsc-SD &   0.325(9)    \\
$    6p_{1/2}  -   6s   $   &   10.2856 &   9.6839  &   6.2\%   &   9.7450  &   SDpT    &   0.63\%  &   SD-SDpT &   9.745(61)   \\
$    6p_{1/2}  -   7s   $   &   9.3594  &   9.1896  &   1.8\%   &   9.2092  &   SDpT    &   0.21\%  &   SD-SDpT &   9.209(20)   \\
$    6p_{1/2}  -   8s   $   &   1.9219  &   1.8532  &   3.7\%   &   1.8616  &   SDpT    &   0.45\%  &   SD-SDpT &   1.862(8)    \\
$    6p_{1/2}  -   9s   $   &   0.9702  &   0.9364  &   3.6\%   &   0.9364  &   SD  &   0.50\%  &   0.5\%   &   0.936(5)    \\
[0.3pc]
$    6p_{3/2}  -    5s      $   &   0.6055  &   0.5409  &   11.9\%  &   0.5276  &   SDsc    &   2.51\%  &   SDsc-SD &   0.528(13)   \\
$    6p_{3/2}  -    6s      $   &   14.4575 &   13.5918 &   6.4\%   &   13.6804 &   SDpT    &   0.65\%  &   SD-SDpT &   13.680(89)  \\
$    6p_{3/2}  -    7s      $   &   13.5514 &   13.3529 &   1.5\%   &   13.3755 &   SDpT    &   0.17\%  &   SD-SDpT &   13.376(23)  \\
$    6p_{3/2}  -    8s      $   &   2.7047  &   2.6001  &   4.0\%   &   2.6129  &   SDpT    &   0.49\%  &   SD-SDpT &   2.613(13)   \\
$    6p_{3/2}  -    9s      $   &   1.3583  &   1.3056  &   4.0\%   &   1.3056  &   SD  &   0.50\%  &   0.5\%   &   1.306(7)    \\
 \end{tabular}
\end{ruledtabular}
\end{table*}

The BBR shift contributes to Rb frequency standard at $10^{-14}$ level (see
\cite{IEEE} for the review of the present status of BBR shift uncertainties for
all atomic clocks). The most recent value of the BBR shift in Rb microwave
frequency standard is accurate to 1\% \cite{BBR-DF}. As a result, ultimate
relative uncertainty induced by the BBR shift in $^{87}$Rb frequency standard
was significantly larger than that of the $^{133}$Cs frequency standard. We
note that we refer to uncertainty of the scalar Stark coefficient. Actual
experimental uncertainty will also include error due to temperature
stabilization. The calculation of Ref.~\cite{BBR-DF} also disagreed with the
old 1975 theoretical calculation of Ref.~\cite{lee} by 2.5\%. As a result, more
accurate calculation of Rb BBR shift is in order.

In this work,  we  calculated the blackbody radiation shift of the ground-state
hyperfine microwave transition in $^{87}$Rb using the relativistic all-order
method and evaluated the accuracy of our final value.  Our predicted value of
the scalar Stark coefficient, $k_S=-1.240(4)\times10^{-10}$~Hz/(V/m)$^2$ is
 accurate to 0.3\%. Our calculation reduced the uncertainty in Rb frequency standard
  due to BBR shift to the level of accuracy similar to that of the Cs case.

Another motivation for the present work was to provide a systematic approach to
evaluation of theoretical uncertainty using the calculation of BBR shift in Rb
as an example. Modern applications of theoretical atomic calculations
frequently require some knowledge of the accuracy of theoretical numbers. With
new advances in theoretical methods and in computational power, it is essential
to develop consistent strategies to evaluate the accuracy of theoretical data.
Such evaluations are difficult but very beneficial both to specific
applications and benchmark comparisons of theory and experiment. In this work,
we described the evaluation of uncertainties of the electric-dipole matrix
elements, hyperfine matrix elements, and remainders of various  sums in
sufficient detail to demonstrate specific approaches that were used. The
methods of the uncertainty evaluation outlined in this paper can be used for
various other calculations.

\section{Method}

 The electrical field $E$ radiated by a blackbody at temperature $T$
and described by Planck's law induces a nonresonant perturbation of atomic
transitions  at room temperature~\cite{BBR_Farley}. The average electric field
radiated by a blackbody at temperature $T$ is
\begin{equation}
\langle E^2 \rangle = (831.9 ~\mathrm{V/m})^2 \left(\frac{T(K)}{300}\right)^4.
\end{equation}
 The frequency shift of an atomic state due to such an electrical field
can be related to the static electric-dipole polarizability $\alpha(0)$ by
 \cite{POL-Andrei}
\begin{equation}
 \delta \nu = -\frac{1}{2}(831.9~\mathrm{V/m})^2
\left( \frac{T}{T_0} \right)^4
\alpha(0)\left(1+\epsilon\left(\frac{T}{T_0}\right)^2\right), \label{ee5}
\end{equation}
where  $\epsilon$ is a small dynamic correction due to the frequency
distribution, and $T_0$ is usually taken to be 300$K$. The dynamic correction
$\epsilon$  was evaluated in Ref.~\cite{BBR-DF} and was found to be small,
$\epsilon=0.011$, for Rb microwave frequency standard. Therefore, we do not
recalculate it in this work.

\begin{table*} [ht]
\caption{Selection of the ``best set'' values for diagonal and
  off-diagonal  matrix elements of the magnetic hyperfine operator
  $\mathcal{T}^{(1)}$ in $10^{-8}$~a.u.
 Absolute values of the lowest order DHF, all-order SD, and all-order SDpT
  values  are given in columns 2-4.
   \label{tabb} }
\begin{ruledtabular}
\begin{tabular}{lccccccccc}
\multicolumn{1}{c}{}&
 \multicolumn{1}{c}{DHF}&
  \multicolumn{1}{c}{SD}&
   \multicolumn{1}{c}{SDpT}&
  \multicolumn{1}{c}{Expt.}&
    \multicolumn{1}{c}{Final}&
     \multicolumn{1}{c}{Source}&
       \multicolumn{1}{c}{Unc. (\%)}&
\multicolumn{1}{c}{Unc. source} & \multicolumn{1}{c}{``Best set''}
     \\
 \hline
$5s -  5s $ &   22.0830 &   36.1633 &   34.6801 &   34.6810 &   34.6810 &   Expt.   &   0.00\%  &   Expt.   &   34.681  \\
$5s  - 6s $ &   11.4126 &   17.4008 &   16.8497 &   16.8602 &   16.8602 &   Expt.   &   0.06\%  &   Expt-SDpT   &   16.860(10)  \\
$5s  - 7s $ &   7.3042  &   10.9262 &   10.6061 &   10.6086 &   10.6086 &   Expt.   &   0.02\%  &   Expt-SDpT   &   10.609(2)   \\
$5s  - 8s $ &   5.1907  &   7.6957  &   7.4786  &   7.4855  &   7.4786  &   SDpT    &   0.09\%  &   Expt-SDpT   &   7.479(7)    \\
$5s  - 9s $ &   3.9328  &   5.8004  &   5.6404  &   5.6563  &   5.6404  &   SDpT    &   0.28\%  &   Expt-SDpT   &   5.640(16)   \\
\\[0.3pc]
$5p_{1/2}  -   5p_{1/2}$    &   2.4023  &   4.3197  &   4.1460  &   4.1223  &   4.1223  &   Expt.   &   0.2\%   &   Expt.   &   4.122(8)    \\
$5p_{1/2}   -  6p_{1/2}$    &   1.4218  &   2.4431  &   2.3582  &       &   2.3582  &   SDpT    &   0.6\%   &   from $5p_{1/2}$ &   2.358(14)   \\
$5p_{1/2}   -  7p_{1/2}$    &   0.9681  &   1.6390  &   1.5853  &       &   1.5853  &   SDpT    &   0.6\%   &   from $5p_{1/2}$ &   1.585(10)   \\
$5p_{1/2}   -  5p_{3/2} $   &   0.3835  &   0.3396  &   0.3274  &       &   0.3274  &   SDpT    &   1\% &   from $5p_{3/2}$ &   0.327(3)    \\
$5p_{1/2}   -  6p_{3/2} $   &   0.2273  &   0.1946  &   0.1886  &       &   0.1886  &   SDpT    &   1\% &   from $5p_{3/2}$ &   0.189(2)    \\
$5p_{1/2}   -  7p_{3/2} $   &   0.1550  &   0.1312  &   0.1272  &       &   0.1272  &   SDpT    &   1\% &   from $5p_{3/2}$ &   0.127(1)    \\[0.3pc]
$5p_{3/2}   -   5p_{3/2} $  &   1.3496  &   2.7786  &   2.6682  &   2.7229  &   2.7229  &   Expt.   &   0.065\% &   Expt.   &   2.723(2)    \\
$5p_{3/2}   -   6p_{3/2} $  &   0.8000  &   1.5755  &   1.5212  &       &   1.5483  &   av. SD, SDpT  &   1\% &   from $5p_{3/2}$ &   1.548(15)   \\
$5p_{3/2}   -   7p_{3/2} $  &   0.5453  &   1.0583  &   1.0241  &       &   1.0412  &   av. SD, SDpT  &   1\% &   from $5p_{3/2}$ &   1.041(10)   \\
$5p_{3/2}    -  6p_{1/2}$   &   0.2269  &   0.1905  &   0.1845  &       &   0.1845  &   SDpT    &   1\% &   from $5p_{3/2}$ &   0.185(2)    \\
$5p_{3/2}    -  7p_{1/2}$   &   0.1545  &   0.1275  &   0.1236  &       &   0.1236  &   SDpT    &   1\% &   from $5p_{3/2}$ &   0.124(1)    \\
 \end{tabular}
\end{ruledtabular}
\end{table*}

In the case of the optical transitions, the lowest (second) order
polarizabilities of the clock states are different. In the case of the
ground-state hyperfine  microwave frequency standards, the lowest (second)
order polarizabilities of the clock states are identical and the lowest-order
BBR shift vanishes. Therefore, the Stark shift of the ground state $^{87}$Rb
hyperfine interval ($F=2 - F=1$) is governed by
 the static third-order $F$-dependent polarizability
$\alpha_{F}^{(3)}(0)$.

In this work, we evaluate the scalar Stark coefficient $k_S$,
\begin{equation}
\label{kk}
 k_S=-\frac{1}{2} \left(\alpha_{F=2}^{(3)}(0)-
\alpha_{F=1}^{(3)}(0)\right).
\end{equation}

 The expression for the $\alpha_{F}^{(3)}(0)$ is given by
~\cite{beloy}:
\begin{eqnarray}\label{eq-bbr1}
\label{plrz} \alpha^{(3)}_F(0) & = & \frac{1}{3}\sqrt{(2I)(2I+1)(2I+2)}\left\{
\begin{array}{ccc}
j_v &   I   & F \\
  I   &  j_v & 1
\end{array}
\right\} \times \nonumber \\ && g_I \mu_n
\left(-1\right)^{F+I+j_v}\left(2T+C+R\right),
\end{eqnarray}
where $g_I$ is the nuclear gyromagnetic ratio, $\mu_n$ is the nuclear magneton,
$I=3/2$ is the nuclear spin, and $j_v=1/2$ is the total angular momentum of the
atomic ground state. The $F$-independent sums $T$, $C$, and $R$  for  the
ground state of Rb, $|v\rangle \equiv |5s\rangle$, are given by ~\cite{beloy}:
\begin{eqnarray}
T & = &\sum_{m,n \neq 5s} A_T  \frac{\langle 5s \|D \| m\rangle \langle m \|D
\| n\rangle
\langle n \| \mathcal{T}^{(1)}\| 5s\rangle}{\left( E_m - E_{5s}\right)\left( E_n - E_{5s}\right)}, \label{terms}\\
 C & = &\sum_{m,n \neq 5s}  A_C  \frac{\langle 5s \|D \| m\rangle \langle m \| \mathcal{T}^{(1)}\| n\rangle
\langle n \| D\| 5s\rangle}{\left( E_m - E_{5s}\right)\left( E_n - E_{5s}\right)}, \nonumber\\
R & = & \frac{1}{2} \langle 5s \| \mathcal{T}^{(1)}\| 5s \rangle \left( \sum_{m
\in val} - \sum_{m \in core} \right) \frac{|\langle 5s \|D \| m \rangle
|^2}{\left( E_m - E_{5s}\right)^2}, \nonumber
\end{eqnarray}
  where $\langle i \|D \| j\rangle$ are electric-dipole reduced matrix elements
and $\langle i \| \mathcal{T}^{(1)}\| j\rangle$ are the matrix elements of the
magnetic hyperfine operator $\mathcal{T}^{(1)}$. The quantities
  $A_T$ and $A_C$ are the angular coefficients  given in our case by
\begin{eqnarray*}
A_T &=& \frac{(-1)^{j_m+1/2}}{2}\\
A_C& =& (-1)^{j_m-j_n} \left\{
\begin{array}{ccc}
1 &   1/2   & 1/2 \\
  1   &  j_m & j_n
\end{array}
\right\}.
\end{eqnarray*}

\begin{table*} [ht]
\caption{Absolute values of the electric-dipole reduced matrix elements used in
the calculation of the BBR shift and their uncertainties in atomic units
($ea_0$). \label{tab-e1} }
\begin{ruledtabular}
\begin{tabular}{rrrrrrrr}
\multicolumn{1}{c}{Transition}&
 \multicolumn{1}{c}{$n=5$}&
  \multicolumn{1}{c}{$n=6$}&
   \multicolumn{1}{c}{$n=7$}&
    \multicolumn{1}{c}{$n=8$}&
     \multicolumn{1}{c}{$n=9$}&
       \multicolumn{1}{c}{$n=10$}&
\multicolumn{1}{c}{$n=11$}      \\
       \hline
 $ 5s - np_{1/2}$ &4.231(3) &  0.325(9) & 0.115(3)  &0.060(2) & 0.037(1)  &     0.026(1) &0.020(1) \\
 $ 6s - np_{1/2}$ &4.146(27)&  9.75(6)  & 0.993(7)  &0.388(5) & 0.222(3)  &     0.148(2) &0.109(2) \\
 $ 7s - np_{1/2}$ &0.953(2) &  9.21(2)  & 16.93(9)  &1.856(9) & 0.751(8)  &     0.430(6) &0.289(4)  \\
 $ 8s - np_{1/2}$ &0.502(2) &  1.862(8) & 16.00(2)  &25.9(1)  & 2.95(2)   &     1.20(1)  &0.69(1) \\
 $ 9s - np_{1/2}$ &0.331(1) &  0.936(5) & 3.00(2)   &24.5(1)  & 36.7(2)   &     4.25(2)  &1.73(2) \\
 $10s - np_{1/2}$ &0.243(1) &  0.607(3) & 1.474(7)  &4.40(2)  & 34.8(2)   &     49.4(2)  &5.78(3) \\
 $11s - np_{1/2}$ &0.189(1) &  0.442(2) & 0.942(5)  &2.12(1)  & 6.05(3)   &     46.8(2)  &63.9(3) \\[0.5pc]
 $ 5s - np_{3/2}$ &5.978(5) &   0.528(13)& 0.202(4) & 0.111(3) &0.073(2) & 0.053(2)&   0.040(2) \\
 $ 6s - np_{3/2}$ &6.047(34)&   13.68(9) & 1.53(1) & 0.621(7) &0.363(5)  & 0.246(4)&   0.182(3) \\
 $ 7s - np_{3/2}$ &1.350(2) &   13.38(2) & 23.7(1) & 2.82(2)  &1.18(1)   & 0.68(1) &   0.465(7) \\
 $ 8s - np_{3/2}$ &0.708(2) &   2.61(1)  & 23.19(2) & 36.3(2)  &4.45(2)   & 1.85(2) &   1.08(2) \\
 $ 9s - np_{3/2}$ &0.466(2) &   1.306(7) & 4.19(2)  & 35.5(2)  &51.2(3)   & 6.39(3) &   2.66(3) \\
 $10s - np_{3/2}$ &0.341(1) &   0.845(4) & 2.04(1) & 6.13(3)  &50.3(3)   & 68.9(3) &   8.65(4) \\
 $11s - np_{3/2}$ &0.266(1) &   0.614(3) & 1.302(7) & 2.92(2)  &8.40(4)   &  67.7(3)&
 89.2(4)\\
 \end{tabular}
\end{ruledtabular}
\end{table*}

The sums are made finite with the use of finite B-spline basis set in a
  spherical cavity. The sum over the complete
  finite basis set is equivalent to the sum over the
bound states and integration over the continuum. We use a complete set of DHF
wave functions on a
  nonlinear grid generated using B-splines
constrained to a spherical cavity.  A cavity radius of 220~$a_0$ is chosen to
accommodate all $ns$ and $np$ valence orbitals up to $n=12$. The basis set
consists of 70 splines of order 11 for each value of the relativistic angular
quantum number $\kappa$.

 Sums over $m$ and $n$ run over all possible states
allowed
  by the selection rules and limits of the sums.
  Therefore, three distinct sets of matrix elements are
  needed for the present calculations:
 electric-dipole matrix elements between $ns$ and $mp_j$ states, $\langle mp_j \|D \| ns\rangle$, and diagonal and
  off-diagonal  matrix elements of the magnetic hyperfine operator for both $ns$ and $np$ states:
  $\langle ns \|\mathcal{T}^{(1)}\| 5s\rangle$,
   $\langle mp_{j_1} \| \mathcal{T}^{(1)}\| np_{j_2}\rangle$.
 Therefore, the calculation of the BBR shift reduces to the evaluation of the electric-dipole
  and magnetic hyperfine matrix elements.

In this work, we use atomic units, in which, $e$, $m_e$, $4 \pi \epsilon_0$ and
the reduced Planck constant $\hbar$ have the numerical value $1$.
Polarizability in a.u. has the dimension of volume, and its numerical values
presented here are thus expressed in units of $a_0 ^3$, where $a_0 \approx
0.052918$ nm is the Bohr radius. The atomic units for $\alpha$ can be converted
to SI units via $\alpha / h [$Hz$/(V/m)^2]= 2.48832 \times 10 ^{-8} \alpha
[a.u.]$, where the conversion coefficient is $4 \pi \epsilon_0 a_0^3 /h$ and
Planck constant $h$ is factored out.

 We start
our calculation by evaluating all three terms in Dirac-Hartree-Fock (DHF)
approximation. The resulting DHF values
  for the $T$, $C$, and $R$ terms in atomic units are
\begin{eqnarray*}
2T^{\mathrm{DHF}} &=&2.376\times 10^{-3},\quad C^{\mathrm{DHF}}=6.111\times
10^{-6},\quad   \nonumber\\
R^{\mathrm{DHF}} &=&3.199\times 10^{-3}.
\end{eqnarray*}
Then, we replace all dominant matrix elements by the ``best set'' values that
have been evaluated for their accuracy and replace corresponding energies by
their experimental values \cite{nist-web,Moore}. We refer to the terms where
such replacements have been made as ``main'' terms, and refer to the remaining
terms calculated in the DHF approximation as remainders.

We note that it is essential not to mix DHF and high-precision data within a
single contribution. For example, experimental energies should not be combined
with DHF matrix elements in any of the terms. In the present calculations, all
data in main terms are high-precision theory or experiment values and all data
in the remainders and in core terms are taken to be DHF values. Mixing values
of significantly different accuracy leads to fictitious changes in the final
results, in particularly in Term T. We carried out numerical tests that support
this statement, and we attribute this issue to the violation of the finite
basis set completeness.

 We note that while we use the experimental values of
the energies in the main terms, the accuracy of our all-order theoretical
energy values is very high. We made extensive comparison of removal energies
calculated using the SD all-order method and experimental values
\cite{nist-web,Moore} for the $(5-11)s$, $(5-10)p_j$, $(4-10)d_j$, and
$(4-7)f_j$ states.  Additional first and second-order Breit contributions, Lamb
shift, and third-order Coulomb correlation correction not accounted for by the
SD approximations were also included into the calculation. Our values agree to
better than 10~cm$^{-1}$ for all levels with exception of the $5s$, $6s$, $7s$,
$4d_{3/2}$, and $4d_{5/2}$ levels, where the differences are 27, 25, 12, 32,
and 29~cm$^{-1}$, respectively. We note that the ground state energy is
-33691~cm$^{-1}$, making the agreement better than 0.1\%.

\section{''Best set'' matrix elements and their uncertainties}
 The ``best
set'' consists of our all-order high-precision results  and several
experimental values. The following 128  matrix elements have been replaced by
the all-order or experimental values:
\begin{eqnarray*}
&&\langle mp_j \|D \| ns\rangle, \, m=5-12, \,n=5-12;\hspace{1cm}\\
&&\langle ns \|\mathcal{T}^{(1)}\| 5s\rangle, \,n=5-12; \\
&&\langle mp_{j_1} \| \mathcal{T}^{(1)}\| np_{j_2}\rangle, \,m=5-7,\,  n=5-7.
\end{eqnarray*}
The all-order calculation of Rb matrix elements has been described in detail in
\cite{rb}.

We illustrate the selection of the ``best set'' values of the electric-dipole
matrix elements and determination of their uncertainties in Table~\ref{taba-1},
where we list a few examples. The complete table is given in
Ref.~\cite{IEEE2010}. The absolute values in atomic units ($ea_0$) are given in
all cases. We list the lowest-order DHF results, all-order SD values, and their
relative differences in percent in columns 2 - 4 of Table~\ref{taba-1}. The
relative differences of DHF and single-double (SD) all-order numbers give a
good estimate of the size of the correlation correction. In general, the
smaller the correlation correction, the more precise our theoretical values
are.
 The
final values used in our ``best set'' are listed in column 5. The next column
identifies the source of these values for each of the matrix elements. The
$5s-5p_j$ matrix elements are experimental values from \cite{volz}. All other
E1 matrix elements are from all-order calculation that included SD, SDpT, or
SD$_\textrm{sc}$ values. The SD$_{sc}$ values include additional corrections
added to SD \textit{ab initio} results by means of the scaling procedure
described in Ref.~\cite{review07} and references therein. The SDpT label refers
to \textit{ab initio} all-order calculations that include single, double, and
partial triple excitations. The selection of the particular value as final is
determined by the study of the dominant correlation correction terms (because
the scaling procedure is only applicable for certain classes of terms) and
accuracy requirements. In the present calculation, very high accuracy is not
needed for matrix elements with high values of principal quantum numbers. In
such cases, SD values are sufficiently accurate for E1 matrix elements.

 Evaluation of theoretical uncertainties is a very difficult
problem since it essentially involves evaluation of the quantity that is not
known beforehand. Several strategies can be used in evaluating the
uncertainties of the all-order results, including the study of the breakdown of
the various all-order contributions, identification of the most important
terms, and semi-empirical determination of  important missing contributions.
Our uncertainty estimates are listed in percent in column labeled ``Unc.''. The
method for determining uncertainty is noted in the next column labeled ``Unc.
source''. Where the scaling was performed, it is expected to estimate the
dominant missing correlation correction (see Ref.~\cite{review07} and
references therein for explanation). Therefore, it is reasonable to take the
difference of the \textit{ab initio} and scaled results as the uncertainty.
This is indicated by SDsc-SD note in the ``Unc. source'' column. We note that
this procedure is expected to somewhat overestimate the uncertainty.

In some cases, where such high accuracy was not required but the same
correlation terms were dominant, we carried out \textit{ab initio} SDpT
calculation (i.e. partially included triples) instead and took these values as
final. The uncertainties were estimated  at the differences of the SD and SDpT
numbers in those cases. We note that numerous tests were conducted in the past
that demonstrated that the above-mentioned procedures of the uncertainty
estimates are valid (see Ref.~\cite{review07} for review of the all-order
method and its applications). In the cases of transitions with high values of
the principal quantum numbers (for example, $np - 10s$ transitions) where only
rough estimates of uncertainties were needed, we used uncertainty estimate from
the previous transition. For example, we use 0.5\% as uncertainty estimate for
the $6p_j-9s$ transitions since the uncertainty for the $6p_j-8s$ ones was
0.5\%. Since relative correlation correction generally decreases with $n$, such
procedure can overestimate the uncertainty, but should not underestimate it.
The final results and their uncertainties are summarized in the last columns of
Table~\ref{taba-1}.

The best set values for the electric-dipole matrix elements and their
uncertainties are summarized in Table~\ref{tab-e1}.

 Selection of the ``best set'' values for diagonal and
  off-diagonal  matrix elements of the magnetic hyperfine operator
  $\mathcal{T}^{(1)}$ in $10^{-8}$~a.u is illustrated in Table~\ref{tabb}.
  The complete table is given in Ref.~\cite{IEEE2010}.
  To convert the diagonal matrix element in atomic units to
  hyperfine constants in MHz, one multiplies the values in Table~\ref{tabb}
  by
  $$
  \frac{6.5797\times 10^{9}g_I }{\sqrt{j_v(j_v+1)(2j_v+1)}},
  $$
  where the nuclear gyromagnetic ratio $g_I=1.83416
$ for $^{87}$Rb and $j_v$ is total angular momentum of the electronic state.
Triple corrections are large for hyperfine matrix elements and have to be
included. Scaling procedure can not be applied here since the terms that it
corrects are generally not dominant unlike the cases of the $ns-n^{\prime}p$
matrix elements above. The remaining columns in Table~\ref{tabb} are the same
as in the E1 matrix element tables.

\begin{table}
\caption{Absolute values of the diagonal and
  off-diagonal  matrix elements of the magnetic hyperfine operator $\mathcal{T}^{(1)}$ in $10^{-8}$~a.u.
  See text for conversion of
  diagonal matrix elements in atomic units to hyperfine constants in MHz. \label{tab-hyp} }
\begin{ruledtabular}
\begin{tabular}{crcr}
\multicolumn{1}{c}{Matrix element}&
 \multicolumn{1}{c}{Value}&
 \multicolumn{1}{c}{Matrix element}&
 \multicolumn{1}{c}{Value}\\
       \hline
$5s  -  5s  $ & 34.681      &$5p_{1/2} - 5p_{1/2}$& 4.122(8)    \\
$5s  -  6s  $ & 16.86(1)    &$5p_{1/2} - 5p_{3/2}$& 0.327(3)    \\
$5s  -  7s  $ & 10.609(2)   &$5p_{1/2} - 6p_{1/2}$& 2.36(1) \\
$5s  -  8s  $ & 7.479(7)    &$5p_{1/2} - 6p_{3/2}$& 0.189(2)    \\
$5s  -  9s  $ & 5.64(2)     &$5p_{1/2} - 7p_{1/2}$& 1.59(1) \\
$5s  -  10s $ & 4.45(1)     &$5p_{1/2} - 7p_{3/2}$& 0.127(1)    \\
$5s  -  11s $ & 3.63(1)     &                  &\\[0.5pc]
$5p_{3/2} - 5p_{3/2}$&  2.723(2)    &$6p_{1/2} - 6p_{1/2}$& 1.3453(3)\\
$5p_{3/2} - 6p_{1/2}$&  0.185(2)    &$6p_{1/2} - 6p_{3/2}$& 0.108(1)\\
$5p_{3/2} - 6p_{3/2}$&  1.55(2) &$6p_{1/2} - 7p_{1/2}$& 0.902(2)\\
$5p_{3/2} - 7p_{1/2}$&  0.124(1)    &$6p_{1/2} - 7p_{3/2}$& 0.073(1)    \\
$5p_{3/2} - 7p_{3/2}$&  1.04(1) &       &\\ [0.5pc]
$6p_{3/2} - 6p_{3/2}$&  0.889(1)    &$7p_{1/2} - 7p_{1/2}$& 0.6020(3)\\
$6p_{3/2} - 7p_{1/2}$&  0.072(2)&    $7p_{1/2} - 7p_{3/2}$&  0.049(1)\\
$6p_{3/2} - 7p_{3/2}$&  0.58(1) &$7p_{3/2} - 7p_{3/2}$&  0.4034(3)\\
 \end{tabular}
\end{ruledtabular}
\end{table}

Most of the diagonal hyperfine matrix elements are taken from the experiment.
Experimental uncertainties are listed where experimental data are used.
 Off-diagonal hyperfine matrix elements between the $s$-states
  $\langle ns \| \mathcal{T}^{(1)}\| n^{\prime}s\rangle$ can be also
  evaluated from experimental hyperfine constants using the formula
  \begin{equation}
   |\langle ns \| \mathcal{T}^{(1)}\| n^{\prime}s\rangle| = \sqrt{\langle ns \| \mathcal{T}^{(1)}\| ns\rangle
   \langle n^{\prime}s \| \mathcal{T}^{(1)}\| n^{\prime}s\rangle},
  \end{equation}
  that is useful for the cases where accurate values of the hyperfine constants $A$ are available.
We list such values for the off-diagonal matrix elements as experimental.
 Since a
large number of high-precision experimental values is available for matrix
elements in Table~\ref{tabb}, the remaining uncertainties for off-diagonal
matrix elements are assigned based on the differences of the theory values for
the most relevant diagonal matrix elements with experiment. For example, the
entry ``from $5p_{1/2}$'' in the ``Unc. source'' column indicates that the
difference of the theoretical $5p_{1/2}$ hyperfine constant with the
experimental value was used to assign the uncertainty of the off-diagonal
matrix element.  We note that contributions of the $np-np^{\prime}$ matrix
elements to total uncertainty of the static Stark coefficient $k_S$ is very
small, and approximate estimate of uncertainties is sufficient.

The best set values for the hyperfine matrix elements and their uncertainties
are summarized in Table~\ref{tab-hyp}.

\section{BBR shift uncertainty}
The total uncertainty of the main terms of the static Stark coefficient is
obtained by adding uncertainties from all contibutions in quadrature. The
uncertainties in the remainders are evaluated separately for each term.

\begin{table}
\caption{ Comparison of the DHF values for the main contributions $\left(
\sum_{m=5}^{12}\right)$ to term T with the final ``best set'' values. $n$
refers to terms of the $ns$ sum. The relative difference between the two values
is given in the last column.\label{termT} }
\begin{ruledtabular}
\begin{tabular}{crrr}
\multicolumn{1}{c}{$n$}&
 \multicolumn{1}{c}{DHF}&
 \multicolumn{1}{c}{Final}&
 \multicolumn{1}{c}{Dif.}
 \\
       \hline
  6   &   0.0016114  &   0.0015159(83)   &   -6.3\%  \\
  7   &   0.0002277  &   0.0002156(18)   &   -5.6\%  \\
  8   &   0.0000787  &   0.0000756(7)    &  -4.1\%   \\
  9   &   0.0000378  &   0.0000365(5)    &  -3.7\%   \\
 10   &   0.0000217  &   0.0000209(4)    &  -3.7\%   \\
 11   &   0.0000141  &   0.0000135(4)    &  -3.9\%   \\
 12   &   0.0000104  &   0.0000099(4)    &  -5.0\%   \\
 \end{tabular}
\end{ruledtabular}
\end{table}

Term T contains two sums, over $ns$ and over $mp_{j}$. First, we study the the
remainder of the $mp_j$ sum, ($m>12$) for each of the first few $ns$ terms,
i.e. we break down each $ns$ term as
$$\sum_{ns} \left( \sum_{2p_j}^{12p_j} [...] + \sum_{13p_j}^{Np_j}[...]\right). $$
 There is no $5s$ term
according to Eq.~(\ref{terms}). For $6s$, $7s$, and $8s$ terms, the $m>12$ tail
accounts for only 0.05\%, 0.3\%, and 0.9\%, respectively.  As expected, the
relative tail contribution increases with $n$ since the contributions from
higher $m$ states become relatively more important. However, the contribution
of the $mp_j$ tail is so small for the most important terms that its
uncertainty is negligible. The sum over $ns$ converges much slower, with $n>12$
terms contributing 17\%. Therefore, we had to evaluate the accuracy of the DHF
approximation for the term T. To do so, we used DHF approximation for main $ns$
terms,  and compared the results with out final ``best set'' values. The
comparison is illustrated in Table~\ref{termT}. The columns 2 and 3 contain
main T terms given by Eq.~(\ref{terms}) for each $ns$, $n = 6 - 12$:
\begin{equation}
\sum_{m=5}^{12} A_T  \frac{\langle 5s \|D \| mp_j\rangle \langle mp_j \|D \|
ns\rangle \langle ns \| \mathcal{T}^{(1)}\| 5s\rangle}{\left( E_{mp} -
E_{5s}\right)\left( E_{ns} - E_{5s}\right)}.
\end{equation}
Column 4 gives the relative differences between DHF and final results. We
expect slightly larger differences for $n=6$ and $n=7$ owing to larger relative
correlation corrections for lower n. Then, the ratio is stable and on the order
of 4\%. Slightly larger ratio for $n=12$ is due to cavity size, i.e. $n=12$
basis set orbitals already slightly differ from true DHF orbitals. We conclude
that the accuracy of the DHF approximation for term T is very high, about 4\%.
Therefore, we adjusted DHF tail for the term T by 4\%. We took 100\% of the
adjustment to be the uncertainty of the  term T remainder.

The DHF value for term C is three orders of magnitude smaller than two other
terms. However, it is necessary to evaluate this term accurately as its final
contribution to the total is 0.5\%. Term C also contains two sums, but terms
with ${m,n}=5-7$ account for 97\% of the total making the uncertainty in the
remainder negligible. In fact, $\{m,n\}=5$ term contributes 89\%. The
interesting feature of term C is a very strong cancelations between individual
contributions leading to change of sign between DHF and final values.  We list
DHF and ``best set'' values for individual contributions to term C in
Table~\ref{termC} to illustrate this cancellation. The terms with $m
\leftrightarrow n$ are the same and are added together.

 Term R is essentially defined by the $n=5$ term, that contributes 99.8\% of
 the total. Therefore, its uncertainty is dominated by the experimental uncertainty of the
 $5s-5p_j$ matrix elements~\cite{volz}. The contribution of the remainder and
 its uncertainty is negligible.

\begin{table}
\caption{ Comparison of the DHF values for the main contributions  to term C
with the final ``best set'' values (a.u.).  \label{termC} }
\begin{ruledtabular}
\begin{tabular}{lrr}
\multicolumn{1}{l}{\{m,n\}}&
 \multicolumn{1}{c}{DHF}&
 \multicolumn{1}{c}{Final}\\
 \hline
$ 5p_{1/2} ~5p_{1/2} $  &  3.96E-05 &   3.74E-05    \\
$ 5p_{1/2} ~5p_{3/2} $  &  3.51E-05 &   1.65E-05    \\
$ 5p_{3/2} ~5p_{3/2} $  & -6.77E-05 &  -7.52E-05    \\
 Total $\{m,n\}=5$   &  \textbf{6.94E-06} &  \textbf{-2.13E-05}    \\  [0.5pc]
$ 5p_{1/2} ~6p_{1/2} $  &  1.89E-06 &   1.74E-06    \\
$ 6p_{1/2} ~5p_{3/2} $  &  8.37E-07 &   3.79E-07    \\
$ 5p_{1/2} ~6p_{3/2} $  &  9.53E-07 &   4.52E-07    \\
$ 5p_{3/2} ~6p_{3/2} $  & -3.68E-06 &  -4.07E-06    \\
Total $\{m,n\}={5,6}$& \textbf{-1.63E-09} &  \textbf{-1.49E-06}    \\[0.5pc]
$ 6p_{1/2} ~6p_{1/2} $ &  2.26E-08 &   2.03E-08    \\
$ 6p_{1/2} ~6p_{3/2} $  &  2.28E-08 &   1.05E-08    \\
$ 6p_{3/2} ~6p_{3/2} $  & -5.02E-08 &  -5.55E-08    \\
 Total $\{m,n\}=6$   & \textbf{-4.76E-09} &  \textbf{-2.47E-08}    \\
 \end{tabular}
\end{ruledtabular}
\end{table}

 The resulting final values
  for the $T$, $C$, and $R$ terms in atomic units are
\begin{eqnarray*}
2T& = &2.247(17)\times 10^{-3},\quad C = -2.385(20)\times
10^{-5},\quad  \nonumber\\
R& = &2.769(2)\times 10^{-3}.
\end{eqnarray*}

We substitute these values into the Eqs.~(\ref{kk},\ref{eq-bbr1}) and multiply
the total by $2.48832 \times 10 ^{-8}$ conversion factor (see paragraph on
atomic units above) to obtain our predicted value of the Stark coefficient,
$k_S=-1.240(4)\times10^{-10}$~Hz/(V/m)$^2$. It is in agreement with the  value
$-1.24\times10^{-10}$~Hz/(V/m)$^2$ of Ref.~\cite{BBR-DF} that was estimated to
be accurate to 1\%. It is also in agreement with measurement
$k_S=-1.23(3)\times10^{-10}$~Hz/(V/m)$^2$ by Mowat~\cite{Mowat}.

We use our value of the scalar Stark shift coefficient to calculate the
quantity $\beta$ defined as
\begin{equation}
\beta = \frac{k_S}{\nu_0} \left( 831.9~\textrm{V/m} \right)^2.
\end{equation}
to be -1.256(4)$\times 10^{-14}$.

\section{Conclusion}
We calculated the scalar Stark coefficient $k_S$ for the $^{87}$Rb microwave
frequency standard and carried out detailed evaluation of the uncertainties of
all its contributions. Our calculation reduced ultimate limit to the
uncertainty of $^{87}$Rb frequency standard due to BBR shift to $4\times
10^{-17}$.

\section*{Acknowledgment}

This work was supported in part by US National Science Foundation Grant  No.\
PHY-07-58088. MSS thanks Joint Quantum Institute, University of Maryland Department
 of Physics and
National Institute of Standards and Technology, College Park,
 for hospitality.

%\bibliography{rb}

\end{document}